\documentclass[12pt, a4paper]{article}

\usepackage{amsmath,amssymb,amsfonts}
\usepackage{graphicx}
\usepackage[colorlinks,hypertexnames=false]{hyperref}
\usepackage{cite}

\begin{document}
\title{Extension of the Standard Model with\\ Chern-Simons type interaction}
\author{Volodymyr Gorkavenko${}^1$,
Ivan Hrynchak${}^{1}$,\\
Oleksandr Khasai${}^2$,
Mariia Tsarenkova${}^{1}$\vspace{0.5em}
\\
${}^1$ \it \small Faculty of Physics, Taras Shevchenko National University of Kyiv,\\
\it \small 64, Volodymyrs'ka str., Kyiv 01601, Ukraine\\
${}^2$ \it \small Bogolyubov Institute for Theoretical Physics, National Academy of Sciences of Ukraine\\
\it \small 14-b, Metrolohichna str., Kyiv 03143, Ukraine
}
\date{}

\maketitle
\setcounter{equation}{0}
\setcounter{page}{1}%

\begin{abstract}
    Extension of the Standard Model with Chern-Simons type interaction contains a new vector massive boson (Chern-Simons boson) that couples to electroweak gauge bosons by the so-called effective Chern-Simons interaction. There is no direct interaction between the Chern-Simons bosons and SM fermions. We consider existing restrictions on the parameters of this SM extension, the effective loop interaction of a new vector boson with SM fermions, and the possibility of the manifestation of the long-lived GeV-scale Chern-Simons bosons in collider experiments.

\end{abstract}


\section{Introduction}

While the Standard Model (SM) has proven remarkably successful in describing collider experiments \cite{Cottingham:2007zz}, there is compelling indirect evidence pointing toward the existence of new physics. Some examples of phenomena that SM can not explain are active neutrino oscillations \cite{Bilenky:1987ty,Strumia:2006db,deSalas:2017kay}, dark matter \cite{Peebles:2013hla,Lukovic:2014vma,Bertone:2016nfn,Cirelli:2024ssz}, and the baryon asymmetry of the Universe \cite{Steigman:1976ev,Riotto:1999yt,Canetti:2012zc}. We can suggest the existence of new particles beyond SM (BSM particles) to solve SM's problems.  However, there is also a possibility that there are new particles unrelated to solving these problems.  

A natural question arises: if there are new particles, why have they not been detected in numerous experiments? There are two possible answers. First, the particles might be too heavy, and current accelerator energies are not sufficient to produce them. In this case, detection would require new, more powerful accelerators such as the FCC \cite{Golling:2016gvc,FCC:2018byv}. Alternatively, the new particles could be light enough to be produced nowadays at the existing accelerators \cite{Gorkavenko:2019nqm,Beacham:2020,Lanfranchi:2020crw,Antel:2023hkf}, but their interactions with SM particles are so feebly that they have not yet been observed. The search for such long-lived particles is already underway in the so-called
 intensity frontier experiments such as MATHUSLA \cite{Curtin:2018mvb}, FACET \cite{Cerci:2021nlb}, FASER \cite{FASER:2018ceo,FASER:2018eoc}, SHiP \cite{Anelli:2015pba,Alekhin:2015byh}, NA62 \cite{Mermod:2017ceo,CortinaGil:2017mqf,Drewes:2018gkc}, DUNE \cite{DUNE:2015lol,DUNE:2020fgq},  LHCb \cite{Gorkavenko:2023nbk}, etc.

To develop the phenomenology of these particles, we need to consider the different types they can be. Namely, they can be scalar \cite{Patt:2006fw,Bezrukov:2009yw,Boiarska:2019jym}, pseudoscalar (axionlike) \cite{Peccei:1977hh, Weinberg:1977ma, Wilczek:1977pj,Choi:2020rgn}, fermion (heavy neutral leptons) \cite{Asaka:2005pn,Asaka:2005an,Bondarenko:2018ptm,Boyarsky:2018tvu}, or vector (dark photons) \cite{Okun:1982xi,Holdom:1985ag,Langacker:2008yv,Ilten:2018crw} particles, see detail, e.g., in reviews  \cite{Curtin:2018mvb,Alekhin:2015byh}. Each of these possibilities results in a theory that introduces different new terms in the SM Lagrangian, often referred to as portals. This work focuses on a portal involving a new massive vector boson with Chern-Simons-like interaction.

 Chern-Simons interactions are known to arise in a variety of theoretical models, including those with extra dimensions and string theory frameworks \cite{Antoniadis:2000ena,Coriano:2005own,Anastasopoulos:2006cz,Harvey:2007ca,Anastasopoulos:2008jt,Kumar:2007zza}.

The idea of creating a Chern-Simons portal is based on the phenomenon of chiral anomaly cancellation in physical theories.  Interest in the Lagrangian terms generated by the chiral anomaly
is explained by the fact that their appearance does not depend on the mass of the particle, which
runs along the internal triangular loop. This means that if it exists
a very heavy particle that cannot be born in a collider
experiments and detected directly, then this heavy particle can manifest itself
by generating the corresponding chiral term of the interaction.
Actually, it is
the only way heavy fermions with a mass that can exceed the capabilities of modern accelerators by many orders of magnitude can manifest themselves in low-energy experiments. For example, if
assume that the capacities of modern accelerators are not sufficient
for the discovery of the heavy $t$-quark, then it would manifest itself as a necessary additional term for cancellation of the chiral
anomalies in SM.

\begin{figure}[t]
 \centering
\includegraphics[width=0.48\textwidth]{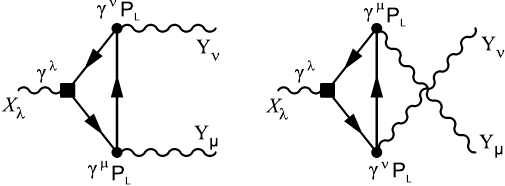}
 \caption{Diagrams generating the Chern-Simons interaction. Heavy fermions, beyond the Standard Model, run in loop triangular diagrams.}
 \label{fig:XAnomaly}
 \end{figure}

Following \cite{Antoniadis:2009ze}, we will be interested in the possibility of the manifestation of the interaction of the new vector massive gauge field $X_\mu$, for example, of the group $U_1(X)$, with SM particles that are generated by the condition of cancellation of the chiral anomaly.
Assume that there are new heavy fermions charged with respect to the gauge group of the SM $U_Y(1)$ and certain additional group $U_X(1)$. At the same time, SM fermions are considered as uncharged with respect to the $U_X(1)$ group and, accordingly, do not directly interact with the $X_\mu$ field. Heavy fermions cannot manifest themselves at modern accelerators; accordingly, it would seem that the $X_\mu$ field cannot manifest itself at low energies either. In \cite{Antoniadis:2009ze}, SM is modified to a theory with symmetry $SU_C(3)\times SU_W(2)\times U_Y (1) \times U_X(1)$, in which the $X_\mu$ boson manifests itself in the interaction with SM vector fields thanks to the diagrams presented in Fig.\ref{fig:XAnomaly}.

As a result of the chiral anomaly effect the interaction between the new vector bosons (henceforth we will call them Chern-Simons or CS bosons) and SM particles is induced in the form of gauge-invariant Lagrangian described by operators with minimum dimension 6 \cite{Alekhin:2015byh,Antoniadis:2009ze}:\newpage

\phantom{bjbjbj}

\vspace{-3.5em}

\begin{align}
    \mathcal{L}_1&=\frac{C_Y}{\Lambda_Y^2}\cdot X_\mu (\mathfrak D_\nu H)^\dagger H B_{\lambda\rho} \cdot\epsilon^{\mu\nu\lambda\rho}+h.c.,\label{L1} \\
    \mathcal{L}_2&=\frac{C_{SU(2)}}{\Lambda_{SU(2)}^2}\cdot X_\mu (\mathfrak D_\nu H)^\dagger F_{\lambda\rho} H\cdot\epsilon^{\mu\nu\lambda\rho}+h.c..\label{L2}  
\end{align}
Here, $\Lambda_Y$ and $\Lambda_{SU(2)}$ represent new scales of the theory, while $C_Y$ and $C_{SU(2)}$ are dimensionless coupling constants. $\epsilon^{\mu\nu\lambda\rho}$ stands for the Levi-Civita tensor ($\epsilon^{0123}=+1$), and $X_\mu$ is a CS vector boson. Please note that $X_\mu$ is a Stueckelberg field \cite{Ruegg:2003ps,Kribs:2022gri}, which ensures the gauge invariance of the Lagrangians \eqref{L1} and \eqref{L2}. The Higgs doublet scalar field is denoted by $H$, and $B_{\mu\nu}=\partial_\mu B_\nu-\partial_\nu B_\mu$ and $F_{\mu\nu}={\displaystyle -ig\sum_{i=1}^3\frac{\tau^i}{2} V^i_{\mu\nu} }$ refer to the field strength tensors of the $U_Y(1)$ and $SU_W(2)$ gauge fields of the SM, respectively.

After electroweak symmetry breaking, the Lagrangians \eqref{L1} and \eqref{L2} produce many terms, including three-field interactions written as four-dimensional operators
\begin{equation}\label{Lcs}  
     \mathcal{L}_{CS}=c_z \epsilon^{\mu\nu\lambda\rho} X_\mu Z_\nu \partial_\lambda Z_\rho +c_\gamma \epsilon^{\mu\nu\lambda\rho} X_\mu Z_\nu \partial_\lambda A_\rho+ \left\{ c_w \epsilon^{\mu\nu\lambda\rho} X_\mu W_\nu^- \partial_\lambda W_\rho^+ + h.c.\right\},
\end{equation}
where $A_\mu$ is the electromagnetic field, and $W^\pm_\mu$ and $Z_\mu$ are the fields associated with weak interactions. The coefficients $c_z$, $c_\gamma$, and $c_w$ are independent and dimensionless. Both $c_z$ and $c_\gamma$ are real, but $c_w$ can be complex $c_w=  \Theta_{W1}+{\rm i} \Theta_{W2}$. Importantly, the CS vector boson $X_\mu$ does not directly interact with SM fermions.

In this paper, we consider existing restrictions on the parameters of this SM extension for the case of light CS bosons, the effective loop interaction of a new vector boson with SM fermions and the possibility of the manifestation of the long-lived GeV-scale CS bosons in collider experiments.

\section{Constrains from interaction with vector\\ fields of SM}

In the case $M_X<M_W, M_Z$ interaction in the form \eqref{Lcs} leads to the additional channel of $W^{\pm}$- and $Z$-boson decay. Namely, the following process are allowed: $Z_\mu\rightarrow X_\mu +\gamma $, $W^\pm_\mu\rightarrow X_\mu+ q_n +{\bar q}_m$,  $W^+_\mu\rightarrow X_\mu+ \ell_n +{\bar \nu}_{\ell n}$.

The corresponding decay width of $Z$-boson can be easily calculated: 
\begin{equation}\label{GZXgammaFull}
    \Gamma(Z\rightarrow X\gamma)= c_\gamma^2\frac{M_Z}{96\pi }  \left(1 - \frac{M_X^2}{M_Z^2}\right)^3 \left(1 + \frac{M_Z^2}{M_X^2}\right).
\end{equation}
 In the limit of  small $M_X$, $M_X/M_Z \ll 1$ we have
\begin{equation}
      \Gamma_{ZX}= c_\gamma^2 \frac{M_Z}{96\pi }  \left(\frac{M_Z}{M_X}\right)^2.
\end{equation}

The corresponding decay width of the $W^\pm$ boson can also be analytically calculated in the approximation of massless quarks and leptons and without taking into account hadronization effects
\begin{equation}\label{Gwud}
\Gamma(W^+\rightarrow Xu\bar{d})=\frac{N_C M^3_W G_F |V_{ud}|^2}{3456\sqrt{2}\pi^3} \left(\Theta^2_{W1}F_1(x)+\Theta^2_{W2}F_2(x)\right),
\end{equation}
where $x=M_W/M_X$, $N_C=3$ -- number of quark colors and $F_1$, $F_2$ are dimensionless functions
\begin{multline*}
F_1(x)= \frac{4}{x^2} -
 6\ln{x}\, (24 - 108 x^2 + 20 x^4 - x^6) -392  + 639 x^2 - 274 x^4 + 23 x^6 +\\ 3 x (14\!-\!x^2) (4\!-\!x^2)^{3/2}\left( \pi\!
    -\!2 \arctan\frac{x (3\!-\! x^2)}{ \sqrt{4\! -\! x^2} (1\!-\!x^2)}  \right),
\end{multline*}
\vspace{-2em}
\begin{multline*}
F_2(x)= 4 + 6 x^6 \ln{x} - 45 x^2 + 18 x^4 + 
 23 x^6 + \\ \frac{3 x^3 (20\! +\! 2 x^2\! -\! x^4)}{\sqrt{4\! -\! x^2}} \left(\pi\! -\! 2 \arctan\frac{x (3\! -\! x^2)}{\sqrt{4\! -\! x^2} (1\! -\! x^2)} \right).
\end{multline*}
It should be noted that in \eqref{Gwud} there is no terms $\sim \Theta_{W1} \Theta_{W2}$. So contributions from real and imaginary parts of coupling $c_w$ do not interfere with each other.

In the limit of  small $M_X$, $M_X/M_W \ll 1$ we have
\begin{equation}
   \Gamma(W^+\rightarrow Xu\bar{d})=\Theta^2_{W1}\frac{N_C M^3_W G_F |V_{ud}|^2}{864\sqrt{2}\pi^3}\left(\frac{M_W}{M_X}\right)^2\!.
\end{equation}

Taking into account channels of decay $W$-boson into $X$-boson and quarks (main contribution is from decay into $u,\bar d$ and $c,\bar s$ because of diagonal elements of the  CKM matrix)  and into three generations of the charged lepton and corresponding neutrino also.  So the full decay width of the $W^+$-boson with the production of $X$-boson for the case $M_X/M_W \ll 1$ can be written as 
\begin{equation}\label{GWXll}
    \Gamma_{WX}= \frac{\Gamma(W^+\rightarrow Xu\bar{d})}{| V_{ud}|^2}\left( |V_{ud}|^2+ |V_{cs}|^2+3/N_C\right) \approx
  \Theta^2_{W1}\frac{M^3_W G_F }{96\sqrt{2}\pi^3}\frac{M_W^2}{M_X^2}.
\end{equation}

Current measurements of the decay width of the $W^\pm$, $Z$ bosons are in agreement with SM predictions. However experimental measurements have some 
uncertainties \cite{ParticleDataGroup:2024cfk}, namely we have 
$\Gamma_W = 2.085 \pm 0.042$ GeV and $\Gamma_Z = 2.4955 \pm 0.0023$ GeV.   
So, new possible channels of the $W^\pm$, $Z$ decay into CS boson can exist  only if $\Gamma_{ZX}<2\Delta\Gamma_Z$ and  $\Gamma_{WX}<2\Delta\Gamma_W$, where $\Delta \Gamma_Z=0.0023$ GeV, $\Delta\Gamma_W=0.042$. Therefore we can 
roughly estimate from above the magnitude of couplings  \cite{Alekhin:2015byh}
\begin{equation}
    c_\gamma^2 \lesssim 10^{-6}\left(\!\!\frac{M_X}{1\,\mbox{GeV} }\!\!\right)^2\!\!, \,\, [Re\, c_w]^2 \lesssim 10^{-2}\left(\!\!\frac{M_X}{1\,\mbox{GeV} }\!\!\right)^2\!\!.
\end{equation}

It should be noted, that in the case of $c_\gamma$ a significantly stronger bound comes from the measuring of the single
photon events at LEP \cite{Acciarri:1997im}.
There the branching at the level $Br=\Gamma_{ZX}/\Gamma_{Z,total} < 10^{-6}$ was established for photons
with  energy above 15 GeV. This leads to the stronger bound $c^2_\gamma 
\lesssim 10^{-9} \left(\frac{M_X}{1 GeV}\right)^2$.  In the case of light CS bosons ($M_X/M_W \ll 1$), we expect $Z\rightarrow X\gamma$ decays with photon energy $E \lesssim  45$ GeV.

\section{Effective loop interaction with fermions of the\\ different flavour}

\begin{figure}[h!]
 \centering
\includegraphics[width=0.45\textwidth]{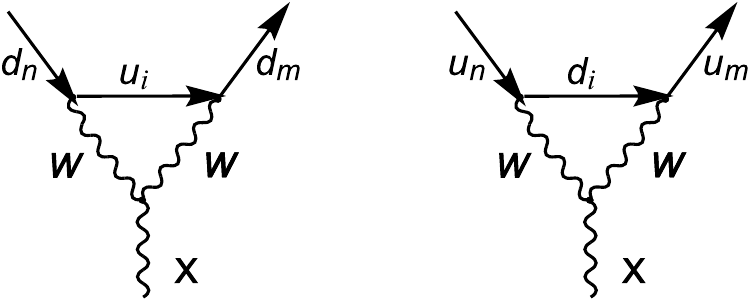}
 \caption{CS boson loop interactions with two quarks of different flavours.}
 \label{fig:loopdiffr}
 \end{figure}

In addition to interaction with SM vector fields, the CS boson can interact effectively with SM fermions due to loop diagrams. 
Such interaction of the CS bosons with quarks of different flavours, see Fig.\ref{fig:loopdiffr},  was considered in \cite{Dror:2017ehi,Dror:2017nsg,Borysenkova:2021ydf}. In this case, effective interaction occurs only due to the interaction of the CS boson with $W^\pm$ boson via coupling $c_w=\Theta_{W1}+\rm{i} \Theta_{W2}$.

It was shown that the divergent part of the loop diagrams is proportional to a non-diagonal element of the unity matrix $V^+V$ ($V$ is the Cabibbo–Kobayashi–Maskawa matrix) and is removed.  It allows one to construct an effective Lagrangian of the interaction of the CS bosons with quarks of different flavours. If $\Theta_{W1}\neq 0$ than the dominant terms of this Lagrangian have form \cite{Borysenkova:2021ydf}
\begin{equation}\label{Mfi5m}
    \mathcal{L}^{CS}_{quarks}=  \sum_{m< n}\Theta_{W1}\left(\! C_{mn}\, \overline{d_m}\, \gamma^{\mu}\,\hat
 P_L  \,  d_n X_{\mu}+C_{nm}^+ \overline{d_n}\, \gamma^{\mu}\,\hat
 P_L  \,  d_m X_{\mu}\!\right),
\end{equation}
where $d_n$ is down-type quark, the summation occurs over the quark generations,
\begin{equation}\label{Cdm}
    C_{mn}= \frac{3a}{2\sqrt{2}\pi^2}\, G_F m_t^2\,V_{d_m t}^+V_{t d_n}\vspace{-0.5em}
\end{equation}
and \vspace{-0.5em}
\begin{align}
& a=0.13,\quad  |C_{sb}| = 1.97\cdot 10^{-4},\nonumber \\
&  |C_{db}|=   4.43\cdot 10^{-5},\quad
   |C_{ds}|=   1.77\cdot 10^{-6}.
\end{align}
As one can see the effective Lagrangian depends only on one of
two  unknown couplings ($\Theta_{W1}$) of the CS boson interaction with $W$ boson.
It was shown that the interaction of the GeV-scale CS boson with up-type quarks can be neglected.

The obtained effective Lagrangian \eqref{Mfi5m} allows us to 
compute dominant production channels of GeV-scale CS bosons in decays of mesons due to reactions $b\rightarrow s + X$, $b\rightarrow d + X$, $s\rightarrow d + X$.
These reactions correspond to the production of the CS bosons in the following decays of charged and neutral mesons:  $B\rightarrow K+X$, $B\rightarrow \pi +X$, $K\rightarrow \pi + X$, where final mesons may also be  
in excited states, see details in \cite{Borysenkova:2021ydf}.

\section{Effective loop interaction with fermions of the\\ same flavour}

\begin{figure}[t]
    \centering
    \includegraphics[width=0.6\textwidth]{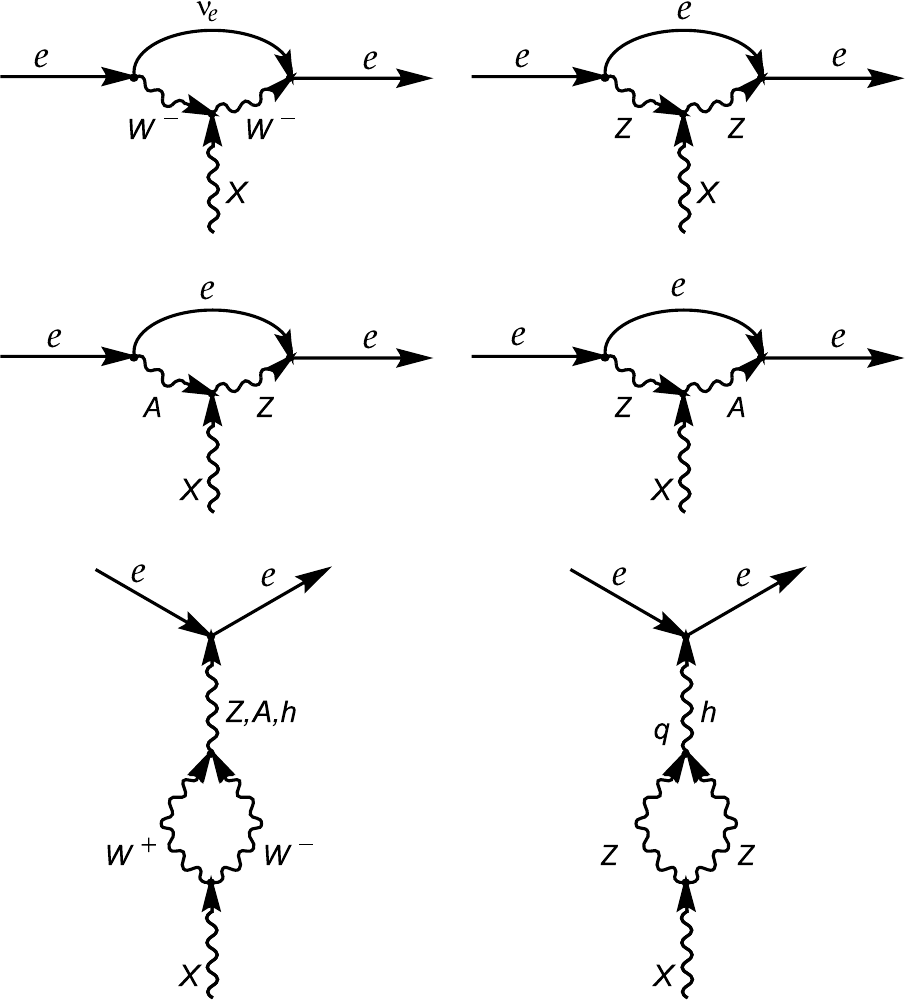}
    \caption{Diagrams of the CS boson's decay into leptons in the unitary gauge. The interaction depends on $c_w$, $c_\gamma$ and $c_Z$ couplings.}
    \label{fig:decay_leptons}
\end{figure}

Unlike the case of effective loop interaction of the CS bosons with quarks of the different flavours, where interaction is defined only by diagrams with $W$-bosons, see Fig.\ref{fig:loopdiffr}  and by coupling $c_w$, in the case of interaction of the CS boson with quarks of the same flavours or with leptons the diagrams with $Z$-bosons and photons are also important, see Fig.\ref{fig:decay_leptons}. Thus the interaction with quarks of the same flavours or leptons depends on $c_w$, $c_\gamma$ and $c_Z$ couplings of Lagrangian \eqref{Lcs}.

We would like the divergences in the loop diagrams of the interaction of the CS bosons with fermions of the same flavours to be automatically removed as well otherwise we will have a problem. These divergences cannot be removed via counterterms because the initial Lagrangian \eqref{Lcs} does not contain terms of the interaction of the CS boson and SM fermions. 

As was shown in \cite{Borysenkova:2021ydf}, for effective loop interaction of CS bosons with quarks of the same flavours or with leptons, divergences in loop diagrams with vertex $XWW$ are not automatically removed during calculations. In \cite{Gor2024CS} this interaction was considered in the unitary gauge taking into account all corresponding diagrams, see Fig.\ref{fig:decay_leptons}. 
It was hoped that the sum of the differences of all diagrams could be cancelled for a certain relation between $c_w$, $c_\gamma$ and $c_Z$ couplings.
Unfortunately, 
it was concluded that using Lagrangian \eqref{Lcs}, we can not eliminate the
divergences in the effective interaction of the CS bosons with fermions of the same flavours. 

We can only hope that, perhaps, further consideration of this problem in non-unitary gauge will help solve the problem of divergences. 
Otherwise, it will mean that the interaction of the CS bosons with fermions of the same flavours
must be considered within the framework of
an effective field theory, namely with the help of corresponding effective operators with a set of new couplings.
Or maybe it will be necessary to supplement the initial Lagrangian \eqref{Lcs} with additional terms, which will allow us to remove differences via corresponding counterterms. 
In both cases, additional couplings appear in addition to $c_w$, $c_\gamma$ and $c_Z$ couplings, which complicates the task of finding manifestations of the CS bosons in experiments.

\section{Discussion}

In this paper, we considered an extension of the SM including a new light massive vector boson with the Chern-Simons interactions. We considered the low energy Lagrangian of the CS boson interaction with vector fields of SM \eqref{Lcs} and we have given the known limits on the parameters of this Lagrangian $c_w$, $c_Z$, $c_\gamma$. Also, we considered the effective loop interaction of the CS bosons with fermions. While the interaction with quarks of different flavours is well defined, the interaction with fermions of the same flavours 
contains divergences that have not yet been removed \cite{Gor2024CS}. 

Even though there is no hope of seeing the direct interaction of the BSM particles with collider detectors, the BSM long-lived particles can still be searched in collider experiments.
The main idea of such experiments is not to directly search for the BSM particles but to search for products of its decay into the SM particles. 
To do this, it is necessary to produce as many BSM particles as possible as a result of SM reactions, e.g., in proton-proton collisions. The produced BSM particles must be further isolated from the SM particles to avoid background events, and then look for very rare decay events of the BSM particles. This is the idea of the intensity frontier experiments mentioned in the Introduction.

As is well known, before the experimental search of a BSM particle at intensity frontier experiments one has to compute the sensitivity region of these experiments, namely 
the parameter region of the new particle  (mass and coupling of interaction with SM particles) where the particle will manifest itself in the experiment. The procedure of computation sensitivity region is well known, see e.g. \cite{Ovchynnikov:2023cry}. It should be noted, that for computation of sensitivity region, one needs to know the technical parameters of the experiment facility and all dominant channels of the BSM particle production and decays.

Let us now consider the possibility of experimentally searching for the CS boson in the intensity frontier experiments.

Regarding production channels, only the production from heavy mesons' decay into the lighter mesons is currently calculated \cite{Dror:2017ehi,Dror:2017nsg,Borysenkova:2021ydf} due to obtained effective Lagrangian \eqref{Mfi5m} of the interaction the CS boson with quarks of different flavours. But the CS boson can be produced and via interaction the CS boson with quarks of the same flavours. For example, in reactions, e.g, $\pi^0\rightarrow X\gamma$, $\omega \to \eta X$, $\phi \to \eta X$ or due to bremsstrahlung, or deep inelastic scattering in proton-proton interactions. 

As for the decay channels, we can only consider decays of the CS bosons into quarks of different flavours followed by hadronization, but even the decays into lepton pairs are not yet available for calculation. 

As one can see, the question of the effective interaction of the CS bosons with fermions of the same flavour is crucial for the experimental search for the Chern-Simons bosons. This is what should be the main focus of further research.\vspace{-0.5em}

\section{Acknowledgments}

The authors would like to thank the Organizers of the conference "New Trends in High-Energy and \mbox{Low-x} Physics" in Sfantu Gheorghe (Romania) for their kind hospitality during this extremely interesting and inspiring meeting.
The work of V.G., I.H., and O.Kh. was supported by the National Research Foundation of Ukraine under project No. 2023.03/0149.

\end{document}